\documentclass[11pt]{article}

\usepackage[final]{acl}

\usepackage{times}
\usepackage{latexsym}

\usepackage[T1]{fontenc}

\usepackage[utf8]{inputenc}

\usepackage{microtype}

\usepackage{inconsolata}

\usepackage{graphicx}
\usepackage{multirow}
\usepackage{booktabs}
\usepackage{placeins}
\usepackage{chngcntr}
\usepackage{tabularx}
\usepackage{float}
\usepackage{pdflscape}
\usepackage{makecell}
\usepackage{array}
\usepackage[table]{xcolor}
\definecolor{revisionblue}{RGB}{0,88,170}
\newif\ifshowrevisions
\showrevisionsfalse
\newcommand{\rev}[1]{\ifshowrevisions\textcolor{revisionblue}{#1}\else#1\fi}

\usepackage{tikz}
\usetikzlibrary{arrows.meta,positioning}
\usetikzlibrary{positioning,fit,arrows.meta}

\usepackage{amsmath}
\usepackage{amssymb}
\usepackage[ruled,vlined]{algorithm2e}
\usepackage{adjustbox}
\usepackage{caption}
\usepackage{enumitem}
\usepackage{float}
\usepackage[T1]{fontenc}
\usepackage{longtable}
\usepackage{linguex}
\usepackage{listings}
\usepackage{multirow}
\usepackage{natbib}
\usepackage{soul}
\usepackage{subcaption}
\usepackage{relsize}
\usepackage{tabularray}
\usepackage{tikz}
\usepackage{comment}

\newcolumntype{Y}{>{\raggedright\arraybackslash}X}
\newcolumntype{L}[1]{>{\raggedright\arraybackslash}p{#1}}

\title{Disentangling the Interpretive and Predictive Roles of LIWC:\\ Controlled Substitution in Depression-Related Classification}

\author{
Hsiang-Chen Yeh \\
Johns Hopkins University \\
\texttt{hyeh10@jhu.edu}
\And
Xiutian Zhao \\
Johns Hopkins University \\
\texttt{xzhao117@jhu.edu}
\And
Aurosweta Mahapatra \\
Johns Hopkins University \\
\texttt{amahapa2@jhu.edu}
\AND
Shreeram Suresh Chandra \\
Johns Hopkins University \\
\texttt{schand30@jhu.edu}
\And
Ryan L. Boyd \\
University of Texas at Dallas \\
\texttt{boyd@utdallas.edu}
\And
Berrak Sisman \\
Johns Hopkins University \\
\texttt{sisman@jhu.edu}
}

\begin{document}
\maketitle

\begin{abstract}
\rev{Linguistic Inquiry and Word Count (LIWC) provides auditable psycholinguistic categories that are widely used to interpret depression-related language, but its incremental predictive role in modern multimodal systems remains unclear. We evaluate LIWC across five English and Chinese depression-related corpora under matched participant-level cross-validation. We ask whether LIWC improves classification and what any performance change reflects. Intact LIWC is compared with three fold-local substitutes: a PCA-rotated version that removes direct access to named category coordinates, a participant-shuffled version that preserves real LIWC profiles while breaking participant alignment, and a random-marginal version that preserves feature-wise distributions. Across multiple fixed representation contexts, the results
provide limited evidence for stable LIWC gains under frozen,
participant-level early fusion. None of the prespecified
dataset-blocked contrasts survives multiple-comparison correction. A separate SBERT calibration produces larger observed intact-versus-shuffled and intact-versus-random separations, indicating that larger participant-aligned signals can produce correspondingly larger separations under the same procedure, while not resolving the five-corpus power limitation. LIWC remains useful as an auditable, corpus-conditioned interpretive layer. These conclusions should not be generalized to fine-tuned, sequence-aware, or end-to-end architectures.}
\end{abstract}

\section{Introduction}

Depression is more than low mood: it can affect emotion, self-understanding, and day-to-day functioning \citep{nolen1991responses, beck2024cognitive}. In language, these changes may surface as distress-related words, heightened self-reference, social withdrawal, or bodily concern. For mental-health NLP, prediction is therefore only one part of the problem. A model may estimate a depression-related label. The harder question is what the verbal pattern means psychologically and which elicitation conditions make it observable.

\rev{Linguistic Inquiry and Word Count (LIWC; \citealp{tausczik2010psychological}) offers one way to make this interpretive question measurable. Throughout the paper, LIWC refers to dictionary-derived category proportions computed from participant-side transcript text.} It maps words onto human-readable categories. These categories include affect and self-reference. They also include social orientation, cognition, temporal focus, and somatic concerns \citep{pennebaker2015development,boyd2022development}. Simply adding LIWC to a depression-related classifier would not be novel on its own. Our question is instead one of attribution: when LIWC changes model performance, what causes that change? One possibility is participant-aligned language information. Another is generic feature structure that resembles LIWC. A third is direct access to the original LIWC category axes. Separating these explanations is the attribution problem we address.

This distinction matters because current depression-detection systems often use speech encoders and transcript embeddings. Many systems also use speech and text fusion or large language models that may already encode affective or semantic information \citep{conneau21_interspeech,reimers2019sentence,zhang-etal-2024-llms,xu2024mental,guo2024large,na2025survey}. A simple add-feature experiment cannot determine whether LIWC contributes participant-aligned signal, nor can it rule out the possibility that any gain merely reflects added dimensions with LIWC-like distributional structure. 

We hypothesize that LIWC's interpretive value and incremental predictive utility are distinct. We therefore evaluate LIWC with matched substitution controls before attributing performance changes to named psychological categories, following broader concerns about faithful interpretation in NLP \citep{jacovi2020towards}. In accordance with prior mental-health NLP work \citep{de2013predicting,coppersmith2014quantifying,shen2022automatic}, we use ``depression detection'' as a conventional task label. Our experiments evaluate corpus-specific depression-related classification, not diagnostic deployment.

We study five English and Chinese depression-related corpora under matched participant-level cross-validation. Because corpora differ in elicitation format, label source, transcript length, language resources, and LIWC coverage, predictive comparisons are conducted within corpus, and cross-corpus LIWC patterns are used descriptively.
More concretely, we frame our inquiry into three research questions:
\begin{enumerate}
    \item Does LIWC add stable predictive value beyond modern speech and transcript representations?
    \item Can LIWC-related performance changes be attributed to participant alignment, generic LIWC-like structure, or access to original LIWC axes?
    \item What corpus-conditioned psychological patterns does LIWC reveal when treated as a descriptive representation rather than a predictive feature?
\end{enumerate}

\paragraph{Contributions.} Our contributions are threefold.
First, we propose a controlled-substitution protocol for evaluating interpretable psycholinguistic feature sets. The protocol compares intact features with fold-local substitutes that preserve different statistical properties, allowing performance changes to be attributed more cautiously to participant alignment, generic feature structure, or access to original interpretable coordinates.
\rev{Second, we instantiate this protocol for LIWC across five English and Chinese depression-related corpora under matched participant-level cross-validation. The primary analysis uses a projected-block route and a WhiSPA-Small route, and a complementary analysis uses a plain XLSR-53 + SBERT route.}
\rev{Third, we find limited evidence for stable LIWC gains under frozen, participant-level early fusion, while the same features support corpus-conditioned descriptive analyses of depression-related language.}

\begin{table*}[t]
\centering
\resizebox{\textwidth}{!}{%
\begin{tabular}{@{}lllrcl@{}}
\toprule
\textbf{Corpus} & \textbf{Language} & \textbf{Elicitation} & \textbf{Size} & \textbf{Class balance} & \textbf{Label / preprocessing note} \\ \midrule
\begin{tabular}[c]{@{}l@{}}DAIC-WOZ\\ \cite{gratch2014distress}\end{tabular} & English           & Virtual interview        & 189           & 56 / 133                          & \begin{tabular}[c]{@{}l@{}}PHQ-8 labels;\\ participant-level aligned features\end{tabular}     \\
\begin{tabular}[c]{@{}l@{}}E-DAIC\\ \cite{ringeval2019avec}\end{tabular}     & English           & Virtual interview        & 275           & 66 / 209                      & \begin{tabular}[c]{@{}l@{}}PHQ-8 labels;\\ participant-text reconstruction\end{tabular}        \\
\begin{tabular}[c]{@{}l@{}}EATD\\ \citep{shen2022automatic}\end{tabular}     & Chinese           & Task-based audio--text   & 162           & 30 / 132                    & \begin{tabular}[c]{@{}l@{}}SDS index-score criterion from\\ corpus definition\end{tabular}     \\
\begin{tabular}[c]{@{}l@{}}MODMA\\ \citep{cai2022multi}\end{tabular}         & Chinese           & Clinical speech          & 52            & 23 / 29                        & \begin{tabular}[c]{@{}l@{}}Diagnosis labels; audited\\  spoken-language subset\end{tabular}    \\
\begin{tabular}[c]{@{}l@{}}PDCH\\ \citep{cao2025multimodal}\end{tabular}     & Chinese           & Psychiatric consultation & 99            & 85 / 14                       & \begin{tabular}[c]{@{}l@{}}HAMD-17 labels; label-valid \\ multimodal intersection\end{tabular} \\ \bottomrule
\end{tabular}%
}
\caption{\rev{Corpus overview for the supervised participant-level experiments. Class balance is positive / negative depression-related labels under each corpus-specific label definition. PHQ-8 denotes the 8-item Patient Health Questionnaire \citep{kroenke2009phq}, SDS denotes the Zung Self-Rating Depression Scale \citep{zung1965self}, and HAMD-17 denotes the 17-item Hamilton Depression Rating Scale \citep{hamilton1960rating}. \rev{Experiments use ten repetitions of stratified 5-fold participant-level cross-validation, yielding 50 held-out evaluations per condition over label-valid feature intersections within each corpus.} Because released challenge splits are pooled and used only as label metadata, the reported values are not comparable to results on the official AVEC test splits.}}
\label{tab:dataset-overview}
\end{table*}

\begin{table*}[t]
\centering
\footnotesize
\setlength{\tabcolsep}{4pt}
\renewcommand{\arraystretch}{1.08}
\begin{tabularx}{0.98\textwidth}{@{}
>{\raggedright\arraybackslash}p{0.17\textwidth}
>{\raggedright\arraybackslash}p{0.43\textwidth}
>{\raggedright\arraybackslash}X
@{}}
\toprule
\textbf{Feature family} & \textbf{Feature settings} & \textbf{Main components} \\
\midrule

\textbf{Speech}
& Single: eGeMAPS \citep{7160715}; Whisper-small \citep{radford2023robust}; XLSR-53 \citep{conneau21_interspeech}; WhiSPA-Small and WhiSPA-Tiny \citep{rao2025whispa}.
  Fusion: eGeMAPS + Whisper; eGeMAPS + XLSR-53; Whisper + XLSR-53; speech fusion; speech fusion + Whisper.
& Acoustic descriptors and frozen speech encoders; pairwise and multi-encoder speech combinations. \\

\midrule

\textbf{Transcript-semantic}
& SBERT transcript semantics \citep{reimers2019sentence}.
& Language-appropriate MPNet sentence embeddings \citep{song2020mpnet} mean-pooled to the participant level. \\

\midrule

\rev{\textbf{Projected block / LIWC}}
& \rev{91-dimensional projected feature block; LIWC; projected block + LIWC.}
& \rev{A learned projected acoustic block and interpretable LIWC category proportions.} \\

\midrule

\rev{\textbf{Text + projected block}}
& \rev{Text + projected-block fusion.}
& \rev{SBERT, the 91-dimensional projected block, and LIWC combined at the participant level.} \\

\midrule

\textbf{Full multimodal}
& Full multimodal fusion; full multimodal fusion + Whisper.
& \rev{Speech representations combined with SBERT, the projected block, and LIWC in early-fusion settings.} \\

\bottomrule
\end{tabularx}
\caption{\rev{Representation contexts used to define the LIWC comparison routes. The table groups the five feature families used to evaluate LIWC in context. Citations identify representation sources, not prior implementations of the present fusion settings. ``+ Whisper'' adds Whisper-small to the named fusion.}}
\label{tab:representation-contexts}
\vspace{-1mm}
\end{table*}

\section{Related Work}

Prior work motivates our study from three connected lines: psycholinguistic dictionary analysis, modern multimodal depression-related NLP, and control-based approaches to interpretability and attribution.

\subsection{LIWC in Depression-Related Language Analysis}
LIWC converts text into human-readable lexical categories with psychological interpretations, and has long been used as a measurement tool in psychological language analysis \citep{pennebaker2003psychological,tausczik2010psychological,boyd2021natural}. In depression-related research, LIWC and related lexical measures have been associated with clinical samples, social-media language, and medical-record-linked outcomes \citep{rude2004language,edwards2017meta,eichstaedt2018facebook}. Reported markers include first-person singular pronouns, negative emotion, absolutist language, cognitive processing, and somatic categories, but their direction and stability vary with genre, sample, and elicitation setting \citep{beech2025using,gu2025linguistic}. This literature supports LIWC as an interpretable description layer, but it does not establish that LIWC provides incremental predictive information once modern speech and transcript representations are present.

\subsection{Modern Depression-Related NLP}
Depression-related NLP has increasingly moved from handcrafted lexical and acoustic features toward frozen speech encoders, transcript embeddings, multimodal fusion, and large language models \citep{conneau21_interspeech,reimers2019sentence,zhang-etal-2024-llms,xu2024mental,guo2024large,na2025survey}. Within this broader shift, LIWC or related linguistic features have been used for speech--text feature comparison, Chinese social-media prediction, cross-cultural spoken-language analysis, student essay analysis, and multimodal remote-interview modeling \citep{morales2016speech,zhang2024natural,amorese2025detecting,abutara2025beyond,jiang2024multimodal}. These studies mainly ask whether linguistic or multimodal features are useful for prediction or description. \rev{Our question is narrower and complementary: when LIWC changes performance after being added to frozen speech and transcript-semantic representations, including WhiSPA as a semantically and psychologically aligned speech representation, what property of LIWC accounts for that change?}

\subsection{Interpretability and Controlled Substitution}
Human-readable feature names do not by themselves provide faithful evidence that a model uses the intended psychological construct. This concern parallels broader work on faithful interpretation in NLP \citep{jacovi2020towards}, and is especially important in mental-health settings where labels, elicitation formats, languages, and sampled populations vary across corpora \citep{ernala2019methodological,chancellor2020methods}. A LIWC-related performance gain may reflect participant-aligned lexical signal, but it may also reflect dimensionality, distributional regularities, LIWC-like profile structure, or interactions with the downstream classifier.

Our approach follows the logic of control-based attribution, including permutation-style importance estimates and control-task baselines that test whether a feature set or probe captures the intended signal rather than incidental structure \citep{breiman2001random,hewitt2019designing}. We adapt this logic to interpretable psycholinguistic features by comparing intact LIWC with fold-local PCA-transformed, participant-shuffled, and random-marginal substitutes. This controlled substitution framework treats LIWC as an instance of a broader problem: interpretable feature sets are not self-validating explanations, and predictive gains require matched controls before they can be attributed to named psychological categories.

\section{Datasets and Representations}

Before interpreting the LIWC controls, we describe the corpora that define the participant-level labels and feature intersections, then the representation families used as comparison contexts.

\subsection{Corpora and Preprocessing}

The study uses five depression-related corpora in English and Chinese. They come from different elicitation and clinical settings, including virtual interviews, task-based recordings, clinical speech, and psychiatric consultations. Table~\ref{tab:dataset-overview} reports the participant-level intersections between available labels and extracted features. \rev{Class balance varies substantially across datasets, with PDCH showing the strongest imbalance; macro-F1 is therefore the primary classification metric.} \rev{Released challenge splits are used only as label metadata sources where applicable; all experiments use pooled label-valid feature intersections under participant-level cross-validation. Consequently, the reported E-DAIC results are not directly comparable to prior results evaluated on the official AVEC train, development, and test splits.} Appendix~\ref{app:dataset_details} gives the dataset descriptions, operational label rules, split-pooling decisions, preprocessing details, and LIWC coverage audits.
Because label sources differ across corpora (PHQ-8, SDS, diagnosis, and HAMD-17), the task is corpus-specific depression-related classification rather than measurement of a single unified clinical construct.

\subsection{Feature Representations}
\label{sec:representation-contexts}

\rev{We organize the feature space into representation families that define the modeling contexts for evaluating LIWC. The speech context includes hand-crafted acoustic descriptors, frozen self-supervised speech encoders, and WhiSPA representations. The transcript-semantic context uses language-appropriate MPNet sentence embeddings through SBERT \citep{reimers2019sentence,song2020mpnet}, mean-pooled to the participant level. A separate 91-dimensional projected feature block is used in one comparison route. This block is produced from XLSR-53 features through a learned 1024--256--91 projector and stored internally as \texttt{psycemb\_0}--\texttt{psycemb\_90}. It is not the ten-dimensional PsychEmb teacher target defined by \citet{rao2025whispa}. LIWC remains the only dictionary-category feature set manipulated in the controlled substitution analysis. Combined contexts concatenate text, projected, speech, and multimodal feature groups. Table~\ref{tab:representation-contexts} summarizes these contexts. All representations are converted to participant-level vectors before supervised learning; the fold-local preprocessing and evaluation protocol are described in Section~\ref{sec:eval}.}

LIWC-based predictive comparisons are conducted within corpus. English LIWC-22 and Simplified Chinese LIWC differ in category inventories, segmentation assumptions, and coverage, and language is partly confounded with elicitation and clinical setting in this collection. We therefore use cross-corpus LIWC patterns only as corpus-conditioned descriptions; coverage details are reported in Appendix~\ref{app:label_rules}.

\section{Controlled Substitution Framework}
\label{sec:methodology}

\begin{figure*}[t]
\centering
\ifshowrevisions\color{revisionblue}\fi
\begin{tikzpicture}[
  node distance=6mm and 7mm,
  box/.style={
    draw,
    rounded corners=2mm,
    fill=blue!12,
    text=black,
    align=center,
    minimum height=18mm,
    text width=0.145\textwidth,
    inner sep=4pt,
    font=\scriptsize
  },
  boxwide/.style={
    box,
    text width=0.18\textwidth,
    minimum height=24mm
  },
  toplabel/.style={
    font=\bfseries\scriptsize,
    align=center,
    text=black,
    text width=0.17\textwidth
  },
  arrow/.style={-{Latex[length=2mm]}, thick}
]

\node[box] (data) {participant audio\\and transcript};
\node[toplabel, above=2mm of data] {Five corpora};

\node[box, right=of data] (features) {XLSR-53, SBERT,\\WhiSPA-Small,\\91-d projected block,\\and LIWC};
\node[toplabel, above=2mm of features] {Frozen feature extraction};

\node[box, right=of features] (pooling) {segment / frame pooling\\participant aggregation};
\node[toplabel, above=2mm of pooling] {Participant-level vectors};

\node[box, right=of pooling] (routes) {plain / projected block / WhiSPA\\intact / PCA / shuffled / random};
\node[toplabel, above=2mm of routes] {Route + LIWC condition};

\node[boxwide, right=of routes] (evaluation) {10 $\times$ 5-fold CV\\matched across conditions\\LR + Linear SVM\\dataset-blocked sign-flip tests\\+ BH correction};
\node[toplabel, above=2mm of evaluation] {Evaluation and inference};

\draw[arrow] (data) -- (features);
\draw[arrow] (features) -- (pooling);
\draw[arrow] (pooling) -- (routes);
\draw[arrow] (routes) -- (evaluation);
\end{tikzpicture}

\caption{\rev{Overview of the analysis pipeline. Participant-side audio and transcript data are converted to frozen participant-level representations. Each fixed route-specific base is combined with one LIWC condition under matched participant-level cross-validation. Primary classifier differences are averaged within dataset before exact sign-flip testing across the five corpus blocks and Benjamini--Hochberg correction.}}
\label{fig:pipeline-overview}
\end{figure*}
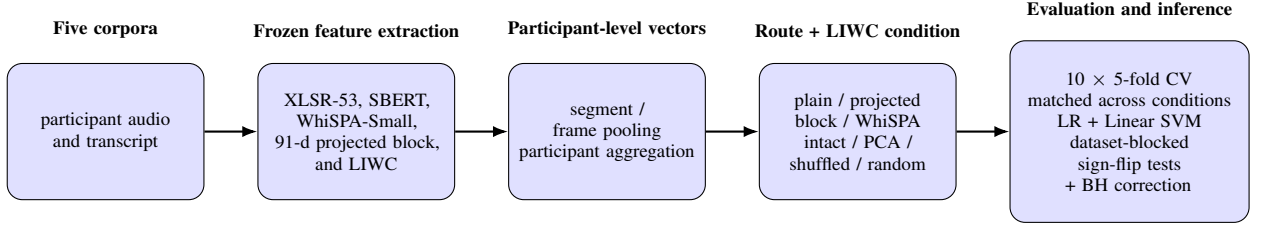

\rev{The framework is organized around one question: when LIWC changes the performance of a modern depression-related classifier, what causes that change? Figure~\ref{fig:pipeline-overview} summarizes the full analysis pipeline. We define three comparison routes in which LIWC is added, then use fold-local LIWC substitutions to test this attribution problem. Finally, we examine LIWC's role as an interpretive representation alongside its controlled predictive contrasts.}

\subsection{Comparison Routes for LIWC Substitution}

\rev{The feature-family comparison is used only to define nontrivial contexts in which LIWC is added; it is not an inferential benchmark of feature families. Throughout the controlled analysis, intact LIWC denotes the original participant-aligned LIWC matrix in its released category basis. Matched LIWC controls denote the fold-local PCA-transformed, participant-shuffled, and random-marginal substitutes generated under the same folds, preprocessing, and classifiers.}

\subsection{Controlled LIWC Substitution}
\label{sec:controlled-substitution}

\begin{figure}[t]
\centering
\scriptsize
\resizebox{0.96\columnwidth}{!}{%
\begin{tikzpicture}[
  font=\footnotesize,
  variant/.style={
    draw,
    rounded corners=2pt,
    semithick,
    fill=white,
    align=center,
    inner xsep=3pt,
    inner ysep=3pt,
    minimum width=24mm,
    minimum height=10mm
  },
  arrow/.style={-{Stealth[length=1.2mm]}, semithick}
]

\definecolor{softbase}{RGB}{248,248,248}
\definecolor{softgroup}{RGB}{246,243,250}
\definecolor{softtable}{RGB}{248,248,248}

\fill[softbase, rounded corners=3pt] (-34mm,-3mm) rectangle (34mm,9mm);
\fill[softgroup, rounded corners=4pt] (-36mm,-47mm) rectangle (36mm,-13mm);
\fill[softtable, rounded corners=3pt] (-34mm,-74mm) rectangle (34mm,-52mm);

\node[align=center] (base) at (0,3mm) {
  \textbf{Route-specific base}\\
  \rev{\textit{Plain:} XLSR-53 + SBERT \quad
  \textit{Projected:} + 91-d block}\\
  \rev{\textit{WhiSPA:} WhiSPA-Small}
};

\node[align=center] (label) at (0,-10mm) {
  \textbf{Augment with one LIWC variant}
};

\draw[arrow] (0,-2mm) -- (label.north);

\draw[semithick] (0,-12.5mm) -- (0,-16mm);
\draw[semithick] (-18mm,-16mm) -- (18mm,-16mm);

\node[variant] (intact) at (-18mm,-25mm) {
  \textbf{Intact}\\
  aligned\\
  original coordinates
};

\node[variant] (pca) at (18mm,-25mm) {
  \textbf{PCA}\\
  fold-local variance\\
  rotated coordinates
};

\node[variant] (shuf) at (-18mm,-40mm) {
  \textbf{Shuffled}\\
  real profiles\\
  wrong participant
};

\node[variant] (rand) at (18mm,-40mm) {
  \textbf{Random}\\
  marginal ranges\\
  no real profile
};

\draw[arrow] (-18mm,-16mm) -- (intact.north);
\draw[arrow] (18mm,-16mm) -- (pca.north);
\draw[arrow] (intact.south) -- (shuf.north);
\draw[arrow] (pca.south) -- (rand.north);

\node[align=center] at (0,-63mm) {
\begin{tabular}{@{}lccc@{}}
\toprule
\textbf{Variant} & \textbf{Align.} & \textbf{Profile} & \textbf{Orig. coord.} \\
\midrule
Intact   & Yes & Yes & Yes \\
PCA      & Yes & Yes & No \\
Shuffled & No  & Yes & Yes \\
Random   & No  & No  & No \\
\bottomrule
\end{tabular}
};

\end{tikzpicture}%
}
\caption{\rev{Controlled LIWC substitution framework. Each fixed route-specific base is augmented with one LIWC condition. The framework tests whether LIWC-related performance changes reflect participant alignment, LIWC-like profile structure, or direct access to the original LIWC axes under the present classifier geometry. The controls are analytical probes; in particular, the PCA contrast should not be interpreted as a test of category semantics alone.}}
\label{fig:liwc-substitution-framework}
\vspace{-1mm}
\end{figure}
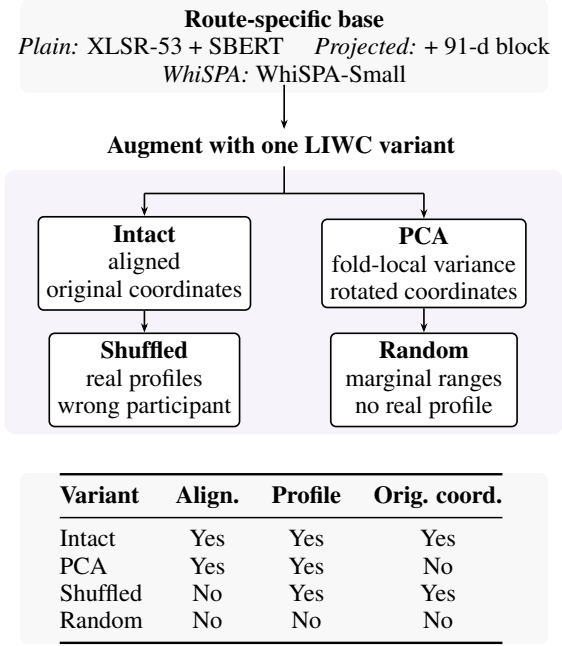

\rev{The controlled LIWC substitution tests are the core experiment of the paper. LIWC is the only feature block whose participant alignment, multivariate profile structure, and access to the original category axes are manipulated. Figure~\ref{fig:liwc-substitution-framework} summarizes the three route-specific bases and the four LIWC conditions. Each route compares intact LIWC with controls generated within the cross-validation loop.}

\rev{We evaluate LIWC in three fixed base contexts. The plain route uses XLSR-53 and SBERT transcript semantics. The projected-block route adds the 91-dimensional projected block described in Section~\ref{sec:representation-contexts} to the same XLSR-53 and SBERT base. Because this block is a learned projection of XLSR-53 features already present in the route, it should not be interpreted as an independent psychological representation. The WhiSPA route uses WhiSPA-Small as the base and tests whether participant-level LIWC adds information beyond a speech representation aligned with semantic and psychological supervision. The projected-block and WhiSPA routes define the eight prespecified primary contrasts. The plain route was added as a complementary analysis and its four contrasts are treated as a separate supplementary family. All route-specific bases are fixed before evaluation and are not selected separately for each dataset.}

PCA-transformed LIWC is generated fold-locally after training-fold imputation and standardization,
then applied to validation LIWC using training-fold parameters. We retain all rank-available
components without whitening and zero-pad to the original width when the fold-local rank is smaller
than the number of LIWC dimensions. This control removes direct coordinate access to the named LIWC
category basis while preserving the training-fold linear subspace up to rank limitations. Consequently,
the intact--PCA contrast should be interpreted narrowly: it tests sensitivity to the original feature
basis under the present preprocessing, rank, classifier, and regularization geometry. It does not by
itself establish that named LIWC categories have unique semantic or psychological predictive content.

Shuffled LIWC preserves real multivariate LIWC profiles while breaking participant alignment. Within each training fold, LIWC rows are permuted among training participants. For validation participants, shuffled controls are sampled from the empirical distribution of training-fold LIWC profiles, so validation-fold LIWC values are not used to construct the control. This procedure preserves realistic LIWC profile structure but decouples profiles from the correct participant identities and labels.

Random LIWC preserves feature-wise empirical marginal distributions while removing both participant alignment and real multivariate profile structure. For each fold, random values are sampled independently from the training-fold empirical marginal distribution of each LIWC feature and then assigned to training and validation participants. For shuffled and random controls, we average performance over 30 independent control draws per fold before cross-dataset paired comparisons. Full control-generation details are provided in Appendix~\ref{app:liwc_control_details}.

\paragraph{Attribution logic.}
We interpret the contrasts jointly. Improvement of intact LIWC over the route-specific base estimates additive value in that representation context. Improvement over PCA-transformed LIWC indicates basis sensitivity under the present fold-local rank, preprocessing, classifier, and regularization setting; it should not be interpreted as a direct test of LIWC category semantics.

\subsection{LIWC Interpretability Analysis}

We also analyze LIWC as an interpretive psychological representation. For visualization, we group selected LIWC categories into eight theory-guided descriptive domains: self-focus, social orientation, negative affect, positive/reward orientation, cognitive processing, somatic/biological concerns, temporal focus, and risk/death language. These domains are descriptive groupings rather than new psychometric scales or official LIWC hierarchies.

\rev{Within each dataset, we compute Cohen's $d$ for each matched LIWC category and average the category-level estimates within domain. Positive values indicate higher category use among participants with depression. Fixed-seed percentile confidence intervals are obtained through participant bootstrap stratified by the depression-related label.} \rev{Domain membership is reported in Table~\ref{tab:liwc-domain-map}; fixed-seed domain-level bootstrap confidence intervals are reported in Tables A12 and A13; and category-level estimates for all mapped categories are reported in Table A14.}

\subsection{Evaluation Protocol}
\label{sec:eval}
Each feature condition is evaluated with eight fixed downstream classifiers, including linear, tree-based, and neural models; full configurations are reported in Appendix~\ref{app:representation_details}. The controlled LIWC substitution tests use logistic regression and Linear SVM as the primary classifiers because the goal is stable matched attribution, not classifier benchmarking. These two models make the LIWC-control contrasts easier to compare under small-sample and imbalanced settings. The remaining classifiers are used as robustness checks over fixed participant-level vectors.

\rev{We convert all features to participant-level vectors before supervised learning. We use ten repetitions of label-stratified 5-fold participant-level cross-validation, yielding 50 held-out evaluations per dataset, classifier, route, and feature condition. Split assignments are matched across feature conditions within each repetition. Training-dependent preprocessing, including imputation, standardization, PCA controls, and stochastic LIWC controls, is fitted or generated using the training portion of each fold only. Classifier and preprocessing configurations are fixed before evaluation for the reported experiments. Neural robustness models use an internal split of the training portion for early stopping; held-out participants are not used for preprocessing, control construction, or model selection. Our claims therefore concern frozen, participant-level, early-fusion representations; sequence-aware, turn-level, and end-to-end models remain outside the present scope.}

\subsection{Statistical Analysis}
\label{sec:statistical-analysis}

\rev{Performance for each dataset, route, classifier, and feature condition is first averaged over the 50 held-out evaluations from ten repetitions of 5-fold participant-level cross-validation. The repeated folds are used to stabilize the within-dataset performance estimates and are not treated as independent inferential units. For the primary cross-corpus LIWC claims, logistic-regression and Linear-SVM mean differences are then averaged within dataset, yielding five dataset-level observations. Two-sided exact sign-flip tests are applied to these five corpus blocks. This avoids treating repeated folds or two classifiers trained on the same participants, labels, and feature constructions as independent cross-corpus evidence.}

\rev{We report a secondary diagnostic analysis over the ten dataset--classifier mean differences. These diagnostic tests assess whether logistic regression and Linear SVM show similar directions, but they are not the primary inferential basis. We report mean differences, bootstrap 95\% confidence intervals, two-sided exact sign-flip $p$-values, and Benjamini--Hochberg adjusted $q$-values. The primary multiplicity family consists of the eight prespecified dataset-blocked macro-F1 contrasts in Table~\ref{tab:main_liwc_substitution_summary}: two additive contrasts and six intact-versus-control contrasts. The complementary plain-route and SBERT calibration analyses each use a separate four-contrast adjustment family and are interpreted as supplementary rather than confirmatory. Other appendix analyses are descriptive. With five dataset-level observations, the smallest possible two-sided exact sign-flip $p$-value is 0.0625. We therefore use bootstrap confidence intervals to summarize effect-size uncertainty, but do not treat interval exclusion of zero as confirmatory evidence unless supported by the exact sign-flip tests and Benjamini--Hochberg adjusted $q$-values.}

\FloatBarrier
\section{Results and Analysis}
\label{sec:results}

\subsection{Comparison Routes Define the LIWC Substitution Context}
\begin{figure}[t]
\centering
\includegraphics[width=\columnwidth]{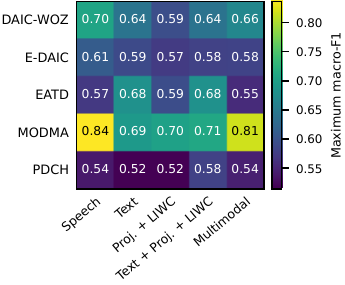}
\caption{\rev{Corpus-dependent upper-envelope macro-F1 by feature family. Each cell is the best macro-F1 across eight classifiers within that family and should be read as an optimistic ceiling, not as average performance. The figure motivates the comparison contexts used in the substitution analysis; it is not a feature-family benchmark. Mean, median, minimum, and maximum summaries across classifiers are reported in Table A4.}}
\label{fig:feature-family-heatmap}
\end{figure}

Before testing LIWC, we summarize the representation landscape in which it is added. Figure~\ref{fig:feature-family-heatmap} reports a descriptive within-family upper envelope across eight classifiers. These optimistic values are not used for inference; they show only that performance varies by corpus and that no single feature family dominates across all datasets. This motivates fixed, nontrivial comparison routes for the LIWC substitution analysis below.

\rev{The three route-specific bases defined in
Section~\ref{sec:controlled-substitution} are fixed in advance rather
than selected separately within each corpus.}

\subsection{Attributing LIWC-Related Performance Changes}
\label{sec:controlled-liwc-results}

We now use the substitution framework to answer the central attribution question: how LIWC-related performance changes should be interpreted once modern representations are already present. The primary LIWC inference is dataset-blocked: logistic-regression and Linear-SVM differences are first averaged within dataset and then compared across the five dataset-level observations. Dataset--classifier contrasts are used only as descriptive diagnostics because the two classifiers share participants, labels, folds, and feature construction. With only five corpus blocks, exact sign-flip tests have coarse resolution, so we interpret the joint pattern of effect sizes, uncertainty intervals, control directions, and Benjamini--Hochberg adjusted $q$-values. Under this resolution, the exact test is used as a conservative directional-consistency check rather than as a test capable of reaching conventional p < .05 significance.

\begin{table}[t]
\centering
\resizebox{\columnwidth}{!}{%
\ifshowrevisions\color{revisionblue}\fi
\begin{tabular}{@{}llrrrr@{}}
\toprule
\textbf{Route} & \textbf{Statistic} & \textbf{Base} & \textbf{PCA} & \textbf{Shuffled} & \textbf{Random} \\
\midrule
\multirow{3}{*}{Projected}
& $\Delta$ F1 & 0.002 & 0.009 & 0.008 & 0.006 \\
& 95\% CI & [-0.004, 0.008] & [-0.001, 0.019] & [0.005, 0.011] & [0.003, 0.010] \\
& $p/q$ & 0.563 / 0.750 & 0.250 / 0.500 & 0.063 / 0.250 & 0.063 / 0.250 \\
\midrule
\multirow{3}{*}{WhiSPA}
& $\Delta$ F1 & -0.008 & 0.024 & -0.001 & -0.003 \\
& 95\% CI & [-0.021, 0.010] & [0.003, 0.056] & [-0.019, 0.021] & [-0.021, 0.022] \\
& $p/q$ & 0.438 / 0.700 & 0.125 / 0.333 & 0.938 / 0.938 & 0.750 / 0.857 \\
\bottomrule
\end{tabular}%
}
\caption{\rev{Main dataset-blocked controlled LIWC substitution contrasts. Columns are comparison conditions; each value is intact LIWC minus the named condition, so positive $\Delta$ F1 favors intact LIWC. Classifier differences are averaged within dataset and compared across five dataset-level observations using two-sided exact sign-flip tests. $q$ values are Benjamini--Hochberg adjusted over the eight prespecified contrasts. Because five-block exact tests have coarse resolution, intervals excluding zero are not treated as multiplicity-adjusted inferential evidence.}}
\label{tab:main_liwc_substitution_summary}
\end{table}

\begin{figure}[t]
\centering
\includegraphics[width=\columnwidth]{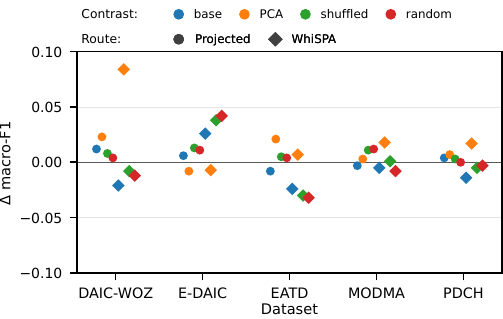}
\caption{\rev{Dataset-level differences from the controlled LIWC substitution analysis. Each marker represents the macro-F1 difference between intact LIWC and the indicated comparison, averaged across logistic regression and Linear SVM within each dataset. Colors denote contrasts, marker shapes denote routes, and the horizontal line indicates zero. The displayed values match those in Table~\ref{tab:dataset-level-liwc-substitution} and are rounded for presentation. Table~\ref{tab:main_liwc_substitution_summary} is computed using the unrounded dataset-level differences. The negative EATD WhiSPA intact-versus-base contrast should be interpreted cautiously in light of EATD's short transcripts and high LIWC sparsity (Table~\ref{tab:liwc-sparsity}).}}
\label{fig:perdataset-dotplot}
\end{figure}

\rev{The main results provide limited evidence for stable LIWC gains under the two prespecified routes. Table~\ref{tab:main_liwc_substitution_summary} reports the dataset-blocked contrasts. Intact LIWC adds $0.002$ macro-F1 over the projected-block base and $-0.008$ over the WhiSPA base. In the projected-block route, intact LIWC is directionally better than the shuffled and random controls by small margins. In the WhiSPA route, the PCA contrast is positive, whereas the shuffled and random contrasts are near zero or slightly negative. None of the eight prespecified dataset-blocked contrasts survives Benjamini--Hochberg correction. The results therefore do not support a stable, practically large cross-corpus LIWC gain or a direct predictive interpretation in terms of named LIWC categories. Bootstrap intervals are treated as descriptive effect-size summaries because the five-block exact tests have coarse resolution.}

\rev{The complementary plain route produces the same general pattern after the 91-dimensional projected block is removed. Intact LIWC adds $0.0027$ macro-F1 over the XLSR-53 + SBERT base, with a 95\% CI of [-0.0064, 0.0104]. The corresponding intact-versus-PCA, shuffled, and random differences are also small. These four contrasts form a separate supplementary family and are reported in Table~\ref{tab:plain-route-supplementary} of Appendix~\ref{app:supplementary_calibration_analyses}.}

\rev{As a calibration, we applied the same substitution procedure to SBERT. The intact-versus-shuffled separation is $0.0820$ [0.0151, 0.1685], and the intact-versus-random separation is $0.0571$ [0.0095, 0.1373]. These observed separations are larger than the LIWC contrasts and provide an effect-size calibration showing that the procedure can yield larger participant-alignment contrasts. None of the four supplementary contrasts survives its separate multiple-comparison correction, and the analysis does not resolve the power limitation of inference based on five corpus blocks. Full results are reported in Table~\ref{tab:sbert-positive-control}.}

\rev{As a supplementary analysis, we predicted LIWC categories from each route-specific base using a five-fold out-of-fold ridge probe with fold-local input preprocessing and fixed $\alpha=1.0$. Mean $R^2$ was negative for every dataset and route, although the proportion of individual categories with $R^2>.10$ varied across corpora and routes. No route was consistently less negative across corpora. The fixed probe did not yield a consistent recoverability pattern and therefore does not explain the opposing directions observed across the projected-block and WhiSPA routes. The route-dependent reversal remains unresolved under the present design; testing whether it reflects representation geometry, regularization, or other interactions requires dedicated analysis. Full settings and results are reported in Table A9.}

\rev{Figure~\ref{fig:perdataset-dotplot} shows the dataset-level contrasts. The direction and magnitude of the LIWC differences vary across corpora, and no dataset shows a consistent advantage across all contrasts and both routes.}

\rev{Table~\ref{tab:diagnostic-liwc-substitution} reports the ten-pair diagnostic estimates, which provide a secondary check rather than the primary inferential basis. They are computed over matched dataset--classifier pairs, whereas Table~\ref{tab:main_liwc_substitution_summary} first averages the two primary classifiers within dataset. In the projected-block route, the diagnostic contrasts favor intact LIWC over PCA, shuffled, and random LIWC by small margins ($\Delta = 0.009$, $0.008$, and $0.006$, respectively). The WhiSPA route shows a positive PCA contrast ($\Delta = 0.024$) and near-zero shuffled and random contrasts ($\Delta = -0.001$ and $-0.003$). None of these diagnostic contrasts survives Benjamini--Hochberg correction. The dataset-blocked and diagnostic estimands therefore support the same qualitative interpretation: any apparent LIWC value is small and route-dependent.}

\subsection{LIWC Contributes Corpus-Conditioned Interpretation}

\begin{figure}[t]
\centering
\includegraphics[width=\columnwidth]{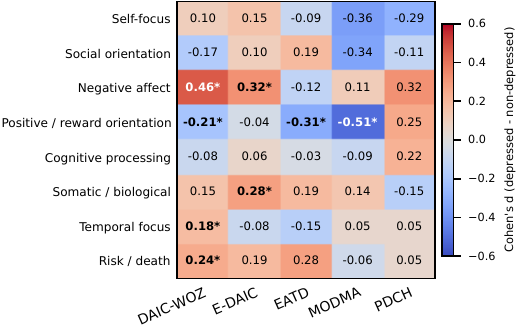}
\caption{\rev{Theory-guided LIWC domain differences. Values are within-dataset Cohen's $d$ estimates; positive values indicate higher category use among participants with depression. Asterisks indicate unadjusted participant-bootstrap 95\% confidence intervals that exclude zero before rounding. They are descriptive and are not corrected for multiple comparisons. LIWC coverage is lower and more variable in the Chinese corpora than in the English interview corpora, so cross-language visual comparisons should be interpreted cautiously. The domains are descriptive groupings rather than official LIWC hierarchies, new psychometric scales, or cross-language equivalent constructs. Domain membership, numeric intervals, and per-category estimates are reported in Appendix~\ref{app:liwc_interpretability_details}.}}
\label{fig:liwc-domain-smd}
\end{figure}

\rev{The substitution tests show limited stable additive predictive value for LIWC, while its human-readable coordinates remain useful as corpus-conditioned interpretive evidence. Figure~\ref{fig:liwc-domain-smd} summarizes LIWC differences between participants with and without depression. \rev{Dictionary coverage is higher in the English interview corpora (DAIC-WOZ: 95.64\%; E-DAIC: 95.63\%) than in the Chinese corpora (EATD: 82.38\%; MODMA: 72.26\%; PDCH: 82.78\%; Table~\ref{tab:transcript_length_coverage}). Cross-language differences in the domain estimates should therefore be interpreted cautiously.} DAIC-WOZ and E-DAIC show similarly positive point estimates for negative affect, somatic or biological language, and risk or death language. This similarity is best read as a shared English virtual-interview pattern with high LIWC coverage and extended participant speech; it should not be taken as evidence that language alone explains the effect.}

\rev{The remaining three corpora show more heterogeneous patterns than the two English interview corpora, consistent with differences in task structure, clinical setting, label source, transcript length, and dictionary coverage. As shown in Table~\ref{tab:liwc-sparsity}, EATD has the shortest participant transcripts in the collection (median 68 tokens) and the highest average LIWC sparsity, with 43.0\% of categories equal to zero per participant. Its risk/death estimate is $0.28$ [-0.16, 0.61], and its self-focus estimate is $-0.09$ [-0.38, 0.23]. These properties may help explain the negative WhiSPA intact-versus-base contrast for EATD, but this remains a post-hoc corpus-specific interpretation rather than a demonstrated causal mechanism. The absence of a stable self-focus pattern is not necessarily inconsistent with prior pronoun findings: elicited interviews and psychiatric consultations may constrain self-reference differently from free writing or social media. These patterns support a situational view of depression-related language, in which psychological constructs become observable differently across elicitation contexts.}

LIWC's value in this analysis is therefore interpretive and bounded. As a dictionary-based summary, it cannot model contextual meaning, negation scope, or turn-level structure. Its role is to provide a psychological vocabulary for describing which aspects of depression-related language become visible in a given corpus, not to define context-free depression markers. Domain definitions and bootstrap uncertainty estimates are reported in Appendix~\ref{app:liwc_interpretability_details}.

\section{Conclusion}
\rev{Across five depression-related corpora, controlled substitutions provide limited evidence for stable LIWC gains under frozen, participant-level early fusion. The projected-block route is compatible with weak participant-aligned lexical information, but no prespecified contrast survives correction. The plain route reproduces the small additive result without the additional projected block, and the SBERT calibration produces larger observed participant-alignment contrasts. These analyses do not establish that LIWC contains no depression-relevant information. They bound the evidence for stable incremental value in the present setting. These conclusions should not be generalized to fine-tuned, sequence-aware, or end-to-end architectures.} \rev{LIWC remains useful as an auditable, corpus-conditioned interpretive layer. However, gains from human-readable features require matched controls before they can be attributed to intended psychological constructs.}

\section*{Limitations}
The corpora are small and heterogeneous in class balance, label source, elicitation format, transcript length, LIWC coverage, segmentation assumptions, and language resources. Accordingly, our LIWC-based predictive comparisons are within-corpus, and cross-corpus LIWC patterns are descriptive rather than cross-lingual psychometric tests. The absence of LIWC effects surviving multiple-comparison correction should be read as limited evidence under this design, not as evidence that LIWC contains no depression-relevant information.

Our conclusions are also bounded by participant-level early fusion. This representation removes turn order, latency, pauses, and temporal change, and should be revisited with turn-level sequence models, late-fusion or gating architectures, fine-tuned encoders, and explicit tests of interactions between LIWC categories and speech or transcript representations. The substitution controls also impose assumptions: PCA is a feature-basis and classifier-geometry control rather than a semantics test; shuffled and random controls rely on finite stochastic draws; and the five-block design has low power. Future work should relax these assumptions using larger corpus collections and finer-grained controls.

LLM-derived psychological annotations could complement LIWC by incorporating contextual meaning, negation scope, and discourse-level structure, but they also introduce prompt sensitivity, model dependence, privacy concerns, and reproducibility challenges. We therefore treat LIWC as a deterministic and auditable baseline for controlled attribution, not as a replacement for LLM-based interpretation.

DAIC-WOZ and E-DAIC share a virtual-interview lineage, and E-DAIC is an extension of the
DAIC-style interview setting. We therefore treat the five dataset blocks as corpus conditions rather
than five fully independent clinical domains. We audited participant identifiers after dataset-specific
normalization and found no overlapping aligned participants between DAIC-WOZ and E-DAIC.
Accordingly, the cross-corpus sign-flip analysis should be read as a conservative descriptive
aggregation over corpus conditions, not as population-level inference over independent clinical
samples.

\section*{Ethical and Release Considerations}
Any future use for individual-level depression screening, diagnosis, or clinical decision support would require external validation, appropriate clinical governance, fairness assessment, privacy safeguards, and clearly defined constraints on use. Potential risks include overgeneralized depression-screening claims, stigmatizing interpretations of language patterns, and misuse of LIWC categories as diagnostic or causal markers. LIWC categories may encode dataset, cultural, or transcription artifacts, and the corpora studied here should not be used to infer individual mental-health status outside their original research context. \rev{The core controlled-substitution implementation, statistical analysis scripts, configuration files, documentation, tests, and a synthetic example are publicly available at \url{https://github.com/jen900704/liwc-controlled-substitution}. Restricted data, participant-derived materials, dataset-specific preprocessing code, and proprietary LIWC resources are not redistributed.} Existing datasets, pretrained models, and proprietary dictionary resources are used under their original access conditions, and derived artifacts are intended for research use only.

\section*{Acknowledgments}
\rev{We thank the Johns Hopkins University Data Science and AI (DSAI) Institute for supporting this research through a faculty startup package.}

\bibliography{custom}
\appendix

\renewcommand{\thetable}{A\arabic{table}}
\renewcommand{\thefigure}{A\arabic{figure}}
\setcounter{table}{0}
\setcounter{figure}{0}

\raggedbottom
\setlength{\textfloatsep}{4pt plus 1pt minus 1pt}
\setlength{\floatsep}{4pt plus 1pt minus 1pt}
\setlength{\intextsep}{4pt plus 1pt minus 1pt}
\setlength{\dbltextfloatsep}{4pt plus 1pt minus 1pt}
\setlength{\dblfloatsep}{4pt plus 1pt minus 1pt}
\setlength{\abovecaptionskip}{2pt}
\setlength{\belowcaptionskip}{0pt}
\renewcommand{\textfraction}{0.05}
\renewcommand{\topfraction}{0.95}
\renewcommand{\dbltopfraction}{0.95}
\renewcommand{\floatpagefraction}{0.85}
\renewcommand{\dblfloatpagefraction}{0.85}
\setcounter{topnumber}{5}
\setcounter{dbltopnumber}{5}
\setcounter{totalnumber}{8}

\newcommand{\apptablesize}{\small}
\newcommand{\appcompactsize}{\scriptsize}
\newcommand{\appstretch}{\renewcommand{\arraystretch}{1.05}}
\newcommand{\appmodelstretch}{\renewcommand{\arraystretch}{0.92}}
\section{Additional Experimental Details}
\label{app:experimental_details}

\subsection{Dataset and Preprocessing Details}
\label{app:dataset_details}

Table~\ref{tab:dataset-overview} in the main text reports the participant-level label--feature intersections used in all supervised experiments. This appendix provides additional dataset and preprocessing details for the five corpora. The final counts reflect the intersection of valid labels and available participant-level feature representations used in the current experiments.

\begin{itemize}
    \item \textbf{DAIC-WOZ} \citep{gratch2014distress,ringeval2017avec} is an English semi-structured clinical interview corpus collected with the virtual interviewer Ellie and designed for research on psychological distress, including depression. We use 189 label--feature-aligned participants with PHQ-8 labels from released AVEC 2017 label metadata pooled for participant-level cross-validation, including 56 positive and 133 negative depression-related labels.
    \item \textbf{E-DAIC} \citep{ringeval2019avec} extends the DAIC-WOZ virtual-interview setting and is used in the AVEC 2019 depression challenge. We use 275 aligned participants, including 66 participants with depression and 209 participants without depression. Because the released transcripts do not always provide clean participant-only text, we reconstruct participant text before extracting transcript-derived features.
    \item \textbf{EATD} \citep{shen2022automatic} is a Chinese emotional audio-textual depression corpus with task-based speech recordings and transcripts. We use 162 aligned participants and apply the SDS index-score criterion from the original corpus definition, yielding 30 participants with depression and 132 participants without depression.
    \item \textbf{MODMA} \citep{cai2022multi} is a Chinese multimodal mental-disorder dataset containing clinically collected speech recordings from patients with depression and matched controls. We use the 52-participant speech subset after metadata auditing, with 23 participants with depression and 29 participants without depression.
    \item \textbf{PDCH} \citep{cao2025multimodal} is a Chinese psychiatric-consultation corpus based on clinical consultations with speech, transcribed text, and HAMD-17 assessments. We use 99 label-valid participants after intersecting clinical labels with available acoustic, semantic, and LIWC features.
\end{itemize}

\paragraph{Operational label rules.}
\label{app:label_rules}

Table~\ref{tab:label-rules} reports the positive-class rule used for each corpus. These rules define corpus-specific depression-related classification targets, not measurements of a single unified clinical construct. All supervised experiments use these binary labels after intersecting valid labels with available participant-level features.

\begin{table*}[!tbp]
\centering
\appcompactsize
\renewcommand{\arraystretch}{1.05}
\setlength{\tabcolsep}{4pt}
\begin{tabularx}{\textwidth}{@{}lXXX@{}}
\toprule
\textbf{Corpus} & \textbf{Positive-class rule} & \textbf{Score field} & \textbf{Label provenance} \\
\midrule
DAIC-WOZ &
PHQ-8 binary label; participants with PHQ-8 score $\geq 10$ are coded positive. &
PHQ8 total or depression severity. &
AVEC 2017 released label metadata pooled for participant-level cross-validation. \\

E-DAIC &
Positive if PHQ-8 total score $\geq 10$; this matches the released binary depression label when available. &
PHQ-8 total / binary depression field. &
AVEC 2019 released detailed label metadata pooled for participant-level cross-validation. \\

EATD &
SDS index score $\geq 53$, with index score defined as raw SDS $\times 1.25$. &
Raw SDS score. &
Corpus-released SDS score metadata. \\

MODMA &
Participant-prefix-derived group label: \texttt{201*} = MDD; \texttt{202*}/\texttt{203*} = healthy control. &
Prefix-derived group. &
Audited participant-prefix metadata. \\

PDCH &
HAMD total score $\geq 8$. &
HAMD total. &
Corpus-released HAMD-17 metadata. \\
\bottomrule
\end{tabularx}
\caption{Operational positive-class rules used in the supervised experiments. Rules follow the corpus-specific label definitions used in the experiments; they are not treated as measurements of a single unified clinical construct.}
\label{tab:label-rules}
\end{table*}
\paragraph{Transcript and LIWC preprocessing audit.}
For each corpus, LIWC features are computed from participant-side transcript text after dataset-specific cleaning in our preprocessing pipeline. Interviewer or agent turns are removed when released speaker labels, transcript structure, or reconstruction rules permit participant-side extraction. E-DAIC participant text is reconstructed before transcript-derived feature extraction because the released transcripts do not always provide clean participant-only text. For the Chinese corpora, participant-side text is exported and analyzed with Simplified Chinese LIWC/SCLIWC resources; tokenization/segmentation and dictionary matching are therefore treated as part of the LIWC software pipeline, not as a separate external preprocessing step. We interpret Chinese LIWC values within dataset because segmentation behavior, dictionary coverage, and corpus domain affect category availability. LIWC coverage is computed after preprocessing as the proportion of analyzed tokens matched by the relevant LIWC dictionary. Missing participant-level features are handled through the label-valid feature intersections reported in Table~\ref{tab:dataset-overview}.

\paragraph{LIWC coverage and cross-language interpretation.}
Because LIWC coverage depends on dictionary version, tokenization or segmentation behavior, transcript length, and corpus domain, we do not interpret Chinese LIWC coefficients as directly comparable to English LIWC-22 coefficients. We treat cross-dataset coefficient patterns involving EATD, MODMA, and PDCH as descriptive. Our coverage audit shows that dictionary coverage differs across languages and corpora. English interview datasets have high mean dictionary coverage (DAIC-WOZ: 95.64\%; E-DAIC: 95.63\%), whereas Chinese datasets show lower and more variable coverage (EATD: 82.38\%; MODMA: 72.26\%; PDCH: 82.78\%). Participant-level transcript-length and coverage statistics are reported in Table~\ref{tab:transcript_length_coverage}. Future work should further examine whether LIWC software settings, segmentation behavior, and corpus-specific transcript quality affect LIWC-derived signals.

\begin{table}[t]
\centering
\appcompactsize
\renewcommand{\arraystretch}{1.05}
\setlength{\tabcolsep}{3.5pt}
\begin{tabular}{lrrrr}
\toprule
\textbf{Dataset} & \textbf{$N$} & \textbf{Mean tokens} & \textbf{Median tokens} & \textbf{Mean cov.} \\
\midrule
DAIC-WOZ & 189 & 1470.85 & 1293.0 & 95.64\% \\
E-DAIC & 275 & 921.17 & 849.0 & 95.63\% \\
EATD & 162 & 96.93 & 68.0 & 82.38\% \\
MODMA & 52 & 695.90 & 641.0 & 72.26\% \\
PDCH & 99 & 1899.49 & 1855.0 & 82.78\% \\
\bottomrule
\end{tabular}
\caption{Participant-level transcript length and LIWC dictionary coverage after dataset-specific preprocessing.}
\label{tab:transcript_length_coverage}
\end{table}

\paragraph{WhiSPA-Small extraction notes.}
For WhiSPA-Small, unreadable audio files are skipped before participant-level pooling. This affected five MODMA files from participant 02010004, but participant-level features were still produced for that participant. PDCH uses the 99-participant label-valid WhiSPA-Small feature set.

\subsection{Representation and Classifier Details}
\label{app:representation_details}

\paragraph{Model sources and versions.}
Table~\ref{tab:model_sources} lists the source and dimensionality of the pretrained representations used in this study. All speech and text encoders are frozen; no encoder weights are updated during supervised training. Acoustic descriptors are extracted with openSMILE eGeMAPS using the Python \texttt{opensmile} package, version 2.6.0. XLSR-53 and Whisper-small are used through Hugging Face \texttt{transformers}, version 4.45.0, with PyTorch 2.5.1. Sentence embeddings are computed with \texttt{sentence-transformers}, version 2.7.0. WhiSPA-Small and WhiSPA-Tiny use the publicly released checkpoints from \citet{rao2025whispa}.

\paragraph{Speech representation extraction.}
For XLSR-53 and Whisper-small, we use the final encoder hidden-state sequence returned by the frozen checkpoint. For WhiSPA-Small and WhiSPA-Tiny, we use the released frozen WhiSPA representation vectors. Segment- or utterance-level vectors are obtained through mean pooling over valid frame positions, then averaged across all retained participant utterances to form one participant-level vector. eGeMAPS descriptors are extracted with openSMILE and aggregated to the participant level using the same participant-level alignment protocol. SBERT sentence embeddings are computed per transcript segment and mean-pooled to the participant level.

\paragraph{Transcript sources.} 
DAIC-WOZ and E-DAIC use released human transcript files, from which we retain participant-side turns using speaker metadata; E-DAIC participant text is reconstructed as described in Appendix~\ref{app:dataset_details}. EATD uses the transcript source released with the corpus. MODMA and PDCH use aligned participant-side transcript resources produced in our preprocessing pipeline; when clean speaker labels are unavailable, diarization and ASR-based reconstruction are used before LIWC and SBERT feature extraction. All LIWC and SBERT features are computed from participant-side text only.

\begin{table}[t]
\centering
\appcompactsize
\renewcommand{\arraystretch}{1.1}
\setlength{\tabcolsep}{3pt}
\begin{tabular}{@{}lll@{}}
\toprule
\textbf{Representation} & \textbf{Source / checkpoint} & \textbf{Dim.} \\
\midrule
\rev{eGeMAPS} & \rev{openSMILE eGeMAPS} & \rev{88} \\
Whisper-small & \texttt{openai/whisper-small} & 768 \\
XLSR-53 & \texttt{facebook/wav2vec2-large-xlsr-53} & 1024 \\
WhiSPA-Small & \citet{rao2025whispa} & 1034 \\
WhiSPA-Tiny & \citet{rao2025whispa} & 394 \\
SBERT & MPNet sentence encoders & 768 \\
\rev{Projected block} & \rev{Learned XLSR-53 projector (1024--256--91)} & \rev{91} \\
LIWC & English LIWC-22 / Simplified Chinese LIWC & corpus-specific \\
\bottomrule
\end{tabular}
\caption{\rev{Representation sources and dimensionalities used in the experiments. All neural encoders are frozen. The 91-dimensional projected block is produced from XLSR-53 features through a learned 1024--256--91 projector; it is not the ten-dimensional PsychEmb teacher target defined by \citet{rao2025whispa}. Dimensionalities are reported before concatenation and participant-level fusion. \rev{Each eGeMAPS block contains 88 acoustic descriptors; identifier and processing-metadata columns are excluded.} LIWC dimensionality is corpus-specific because English LIWC-22 and Simplified Chinese LIWC resources differ in category inventories and retained feature columns.}}
\label{tab:model_sources}
\end{table}

\paragraph{\rev{91-dimensional projected feature block.}}
\rev{The modeling-ready block contains 91 participant-level fields, \texttt{psycemb\_0}--\texttt{psycemb\_90}. The feature-generation pipeline applies a fixed learned 1024--256--91 projector to frozen XLSR-53-derived features, and the resulting modeling-ready files contain one 91-dimensional vector per participant. This internal block is distinct from the ten lexicon-derived PsychEmb teacher dimensions defined in \citet{rao2025whispa}, and the present paper does not assign named psychological meanings to its 91 coordinates.}

\paragraph{Downstream classifiers.}
Each feature condition is evaluated with eight downstream classifiers: logistic regression, Linear SVM, Random Forest, shallow MLP, Deep MLP, Tabular ResNet, Feature-Gated MLP, and 1D-CNN with pooling. The controlled LIWC substitution tests use logistic regression and Linear SVM because they provide stable controlled comparisons under small-sample and imbalanced settings. The broader feature-family analysis uses all eight classifiers as robustness checks across model classes.

\paragraph{Classifier configurations.}
Classifier configurations are fixed before evaluation and applied consistently across datasets and feature conditions. Logistic regression and Linear SVM use balanced class weights. Random Forest uses 300 trees with class-balanced subsampling. Neural classifiers are used as robustness checks over fixed participant-level feature vectors rather than as the inferential basis for the controlled LIWC claims. They are trained for up to 120 epochs with learning rate $10^{-3}$, batch size 16, weight decay $10^{-4}$, and early stopping patience of 20 epochs. Dropout is set to 0.4 for Deep MLP, Feature-Gated MLP, and 1D-CNN with pooling, and to 0.3 for Tabular ResNet. The 1D-CNN is included only as a neural robustness check over fixed participant-level vectors, not as a claim that feature dimensions have a natural temporal or spatial ordering.

\paragraph{Model size and computational budget.}
All pretrained encoders are frozen and used for feature extraction only; no pretrained encoder is fine-tuned. Table~\ref{tab:model_sources} reports the representation dimensionalities used by the downstream classifiers. Experiments were run on institutional CPU/GPU nodes after feature extraction and caching. The main controlled-substitution analyses train small downstream classifiers on participant-level vectors, and stochastic LIWC controls use 30 draws per fold. We did not perform hyperparameter search; classifier configurations are fixed as described above. Exact aggregate GPU hours were not logged, so we report compute qualitatively and will release scripts, configuration files, fold definitions, and seed specifications to support reproducibility.

\paragraph{\rev{Projected-block context.}}
\rev{The 91-dimensional projected block provides little additional value when added to XLSR-53 + SBERT in the feature-family analysis. We retain this route as one fixed comparison context for LIWC substitution and include the plain XLSR-53 + SBERT route as a complementary check. Because the projected block is a learned function of XLSR-53 features already present in the route, it is not interpreted as an independent psychological representation. LIWC remains the named dictionary-category representation used for controlled substitution and interpretation.}

\subsection{Controlled LIWC Substitution Details}
\label{app:liwc_control_details}

All LIWC controls are generated within the cross-validation loop. PCA controls are fitted on training-fold LIWC features and applied to validation-fold LIWC features. Shuffled controls permute training-fold LIWC rows and sample validation controls from the training-fold LIWC distribution. Random controls sample feature values from training-fold LIWC marginal distributions. Control generation uses deterministic fold-specific seeds indexed by dataset, route, control variant, and fold. This procedure makes the substitution results reproducible and prevents validation-fold information from entering control construction. For shuffled and random controls, the main analysis averages performance over 30 independent control draws per fold before cross-dataset paired comparisons are computed. PCA controls remain deterministic because they are fitted from the training-fold LIWC variance structure.

\begingroup
\ifshowrevisions\color{revisionblue}\fi
\begin{table*}[t]
\centering
\scriptsize
\begin{tabular}{llrrrrr}
\toprule
Dataset & Feature family & Mean & Median & Min & Max & $N$ classifiers \\
\midrule
DAIC-WOZ & Speech & 0.615 & 0.640 & 0.472 & 0.702 & 8 \\
DAIC-WOZ & Text & 0.567 & 0.573 & 0.436 & 0.639 & 8 \\
DAIC-WOZ & Projected block / LIWC & 0.556 & 0.568 & 0.469 & 0.592 & 8 \\
DAIC-WOZ & Text + projected block + LIWC & 0.564 & 0.570 & 0.437 & 0.637 & 8 \\
DAIC-WOZ & Multimodal & 0.585 & 0.608 & 0.450 & 0.664 & 8 \\
E-DAIC & Speech & 0.545 & 0.536 & 0.462 & 0.609 & 8 \\
E-DAIC & Text & 0.525 & 0.537 & 0.431 & 0.587 & 8 \\
E-DAIC & Projected block / LIWC & 0.541 & 0.557 & 0.469 & 0.571 & 8 \\
E-DAIC & Text + projected block + LIWC & 0.532 & 0.546 & 0.433 & 0.578 & 8 \\
E-DAIC & Multimodal & 0.528 & 0.536 & 0.433 & 0.579 & 8 \\
EATD & Speech & 0.534 & 0.537 & 0.487 & 0.568 & 8 \\
EATD & Text & 0.574 & 0.568 & 0.487 & 0.678 & 8 \\
EATD & Projected block / LIWC & 0.553 & 0.563 & 0.487 & 0.591 & 8 \\
EATD & Text + projected block + LIWC & 0.587 & 0.590 & 0.487 & 0.678 & 8 \\
EATD & Multimodal & 0.516 & 0.515 & 0.477 & 0.554 & 8 \\
MODMA & Speech & 0.763 & 0.760 & 0.642 & 0.836 & 8 \\
MODMA & Text & 0.638 & 0.648 & 0.532 & 0.689 & 8 \\
MODMA & Projected block / LIWC & 0.656 & 0.663 & 0.581 & 0.699 & 8 \\
MODMA & Text + projected block + LIWC & 0.648 & 0.664 & 0.496 & 0.708 & 8 \\
MODMA & Multimodal & 0.735 & 0.750 & 0.530 & 0.813 & 8 \\
PDCH & Speech & 0.510 & 0.521 & 0.464 & 0.537 & 8 \\
PDCH & Text & 0.459 & 0.466 & 0.398 & 0.515 & 8 \\
PDCH & Projected block / LIWC & 0.484 & 0.481 & 0.458 & 0.516 & 8 \\
PDCH & Text + projected block + LIWC & 0.499 & 0.506 & 0.394 & 0.578 & 8 \\
PDCH & Multimodal & 0.467 & 0.455 & 0.424 & 0.543 & 8 \\
\bottomrule
\end{tabular}
\caption{\rev{Feature-family macro-F1 across eight fixed classifiers. For each dataset and family, each classifier contributes its best available feature setting. The DAIC-WOZ eGeMAPS block contains 88 acoustic descriptors; the non-acoustic \texttt{daic\_shard} processing field is excluded.}}\label{tab:A4-feature-family-no-shard}
\end{table*}

\begin{table*}[t]
\centering
\small
\begin{tabular}{llrrrrr}
\toprule
Route & Contrast & DAIC-WOZ & E-DAIC & EATD & MODMA & PDCH \\
\midrule
Projected & intact-base & +0.012 & +0.006 & -0.008 & -0.003 & +0.004 \\
Projected & intact-PCA & +0.023 & -0.008 & +0.021 & +0.003 & +0.007 \\
Projected & intact-shuffled & +0.008 & +0.013 & +0.005 & +0.011 & +0.003 \\
Projected & intact-random & +0.004 & +0.011 & +0.004 & +0.012 & +0.000 \\
WhiSPA & intact-base & -0.021 & +0.026 & -0.024 & -0.005 & -0.014 \\
WhiSPA & intact-PCA & +0.084 & -0.007 & +0.007 & +0.018 & +0.017 \\
WhiSPA & intact-shuffled & -0.008 & +0.038 & -0.030 & +0.001 & -0.005 \\
WhiSPA & intact-random & -0.012 & +0.042 & -0.032 & -0.008 & -0.003 \\
\bottomrule
\end{tabular}
\caption{\rev{Dataset-level macro-F1 contrasts used as inputs to the dataset-blocked analysis. Each value is intact LIWC minus the named comparison after averaging logistic regression and Linear SVM within dataset.}}
\label{tab:dataset-level-liwc-substitution}
\end{table*}

\begin{table*}[t]
\centering
\begingroup
\ifshowrevisions\color{revisionblue}\fi
\scriptsize
\setlength{\tabcolsep}{4pt}
\begin{tabular}{llrrrrr}
\toprule
\textbf{Route} &
\textbf{Contrast} &
\textbf{$N$} &
\textbf{$\Delta$ macro-F1} &
\textbf{95\% CI} &
\textbf{$p$} &
\textbf{$q$} \\
\midrule
Projected & intact $-$ base     & 10 &  0.0022 & [-0.0034, 0.0075] & 0.4863 & 0.6484 \\
WhiSPA    & intact $-$ base     & 10 & -0.0077 & [-0.0185, 0.0048] & 0.2520 & 0.4031 \\
Projected & intact $-$ PCA      & 10 &  0.0092 & [ 0.0019, 0.0167] & 0.0547 & 0.1094 \\
Projected & intact $-$ shuffled & 10 &  0.0078 & [ 0.0042, 0.0111] & 0.0078 & 0.0625 \\
Projected & intact $-$ random   & 10 &  0.0064 & [ 0.0024, 0.0104] & 0.0254 & 0.0677 \\
WhiSPA    & intact $-$ PCA      & 10 &  0.0239 & [ 0.0064, 0.0452] & 0.0176 & 0.0677 \\
WhiSPA    & intact $-$ shuffled & 10 & -0.0008 & [-0.0139, 0.0137] & 0.9082 & 0.9082 \\
WhiSPA    & intact $-$ random   & 10 & -0.0026 & [-0.0171, 0.0137] & 0.7500 & 0.8571 \\
\bottomrule
\end{tabular}
\caption{\rev{Dataset--classifier diagnostic macro-F1 contrasts over ten matched dataset--classifier pairs. Benjamini--Hochberg correction is applied to these eight tests. These diagnostics are secondary to the five-dataset-block analysis in Table~\ref{tab:main_liwc_substitution_summary}.}}
\label{tab:diagnostic-liwc-substitution}
\endgroup
\end{table*}

\endgroup

\FloatBarrier

\section{Supplementary Analyses and Diagnostics}
\label{app:supplementary_calibration_analyses}

\rev{This appendix reports the complementary plain-route and SBERT calibration contrasts, with Benjamini--Hochberg correction applied separately within each four-contrast family. It also reports a fixed ridge recoverability probe and corpus-level transcript sparsity diagnostics used to contextualize the main findings.}

\begin{table*}[!tbp]
\centering
\begingroup
\ifshowrevisions\color{revisionblue}\fi
\appcompactsize
\begin{tabular}{@{}lrrrr@{}}
\toprule
\textbf{Plain-route contrast} &
\textbf{$\Delta$ macro-F1} &
\textbf{95\% CI} &
\textbf{$p$} &
\textbf{$q$} \\
\midrule
Intact LIWC $-$ base & 0.0027 & [-0.0064, 0.0104] & 0.6250 & 0.6250 \\
Intact $-$ PCA & 0.0111 & [0.0008, 0.0205] & 0.1875 & 0.2500 \\
Intact $-$ shuffled & 0.0085 & [0.0035, 0.0126] & 0.1250 & 0.2500 \\
Intact $-$ random & 0.0081 & [0.0030, 0.0123] & 0.1250 & 0.2500 \\
\bottomrule
\end{tabular}
\caption{\rev{Plain-route contrasts. The base is XLSR-53 + SBERT. Benjamini--Hochberg correction is applied within this separate four-contrast supplementary family.}}
\label{tab:plain-route-supplementary}
\endgroup
\end{table*}

\begin{table*}[!tbp]
\centering
\begingroup
\ifshowrevisions\color{revisionblue}\fi
\appcompactsize
\begin{tabular}{@{}lrrrr@{}}
\toprule
\textbf{SBERT calibration contrast} & \textbf{$\Delta$ macro-F1} & \textbf{95\% CI} & \textbf{$p$} & \textbf{$q$} \\
\midrule
Intact SBERT $-$ base & 0.0455 & [-0.0165, 0.1430] & 0.5625 & 0.5625 \\
Intact $-$ PCA SBERT & 0.0380 & [-0.0035, 0.0975] & 0.3125 & 0.4167 \\
Intact $-$ shuffled SBERT & 0.0820 & [0.0151, 0.1685] & 0.1250 & 0.2500 \\
Intact $-$ random SBERT & 0.0571 & [0.0095, 0.1373] & 0.1250 & 0.2500 \\
\bottomrule
\end{tabular}
\caption{\rev{SBERT calibration analysis using XLSR-53 alone as the base. The same substitution machinery is applied to SBERT. Benjamini--Hochberg correction is applied within this separate four-contrast supplementary family. The analysis provides an effect-size calibration using a representation with larger observed participant-alignment contrasts and does not resolve the low power of five dataset blocks.}}
\label{tab:sbert-positive-control}
\endgroup
\end{table*}

\begin{table*}[t]
\centering
\small
\begin{tabular}{llrrr}
\toprule
Dataset & Route & Mean OOF $R^2$ & Median OOF $R^2$ & Categories $R^2>.10$ \\
\midrule
DAIC-WOZ & Plain & -0.4493 & -0.4807 & 9.65\% \\
DAIC-WOZ & Projected-block & -0.4947 & -0.4841 & 8.77\% \\
DAIC-WOZ & WhiSPA & -0.1658 & -0.1065 & 33.33\% \\
E-DAIC & Plain & -0.9016 & -0.8967 & 0.00\% \\
E-DAIC & Projected-block & -0.8995 & -0.8955 & 0.00\% \\
E-DAIC & WhiSPA & -2.1425 & -2.0987 & 0.00\% \\
EATD & Plain & -1.3036 & -1.1874 & 0.00\% \\
EATD & Projected-block & -1.3028 & -1.1789 & 0.00\% \\
EATD & WhiSPA & -0.4479 & -0.4093 & 7.87\% \\
MODMA & Plain & -0.1629 & -0.1625 & 24.42\% \\
MODMA & Projected-block & -0.1642 & -0.1573 & 24.42\% \\
MODMA & WhiSPA & -0.2548 & -0.2286 & 11.63\% \\
PDCH & Plain & -0.2447 & -0.2452 & 6.59\% \\
PDCH & Projected-block & -0.2471 & -0.2433 & 6.59\% \\
PDCH & WhiSPA & -1.1931 & -1.0649 & 3.30\% \\
\bottomrule
\end{tabular}
\caption{\rev{Direct five-fold out-of-fold recoverability of LIWC categories from the three base routes. Within each fold, missing input features are imputed and input features are standardized using the training data only. Ridge regression uses a fixed penalty of $\alpha=1.0$ without fold-internal hyperparameter tuning; LIWC targets are not standardized. Negative $R^2$ indicates performance below the fold-local mean-prediction baseline. These results characterize linear recoverability under this specified probe rather than an optimized recoverability ceiling. Projected-block values were computed directly and were not inferred from XLSR-53.}}
\label{tab:A9-three-route-recoverability}
\end{table*}

\begin{table*}[!tbp]
\centering
\begingroup
\ifshowrevisions\color{revisionblue}\fi
\appcompactsize
\begin{tabular}{@{}lrrrrr@{}}
\toprule
\textbf{Dataset} & \textbf{Median tokens} & \textbf{LIWC coverage} & \textbf{Categories} & \textbf{Zero-category fraction} & \textbf{Mean nonzero categories} \\
\midrule
DAIC-WOZ & 1293 & 95.64\% & 114 & 0.124 & 99.9 \\
E-DAIC & 849 & 95.63\% & 115 & 0.237 & 87.7 \\
EATD & 68 & 82.38\% & 89 & 0.430 & 50.7 \\
MODMA & 641 & 72.26\% & 86 & 0.120 & 75.7 \\
PDCH & 1855 & 82.78\% & 91 & 0.133 & 78.9 \\
\bottomrule
\end{tabular}
\caption{Transcript length, LIWC coverage, and sparsity. EATD has the shortest transcripts and the highest zero-category fraction. These observations support a post-hoc corpus-specific explanation for unstable LIWC estimates.}
\label{tab:liwc-sparsity}
\endgroup
\end{table*}

\FloatBarrier

\section{LIWC Interpretability Details}
\label{app:liwc_interpretability_details}

\paragraph{Domain definitions and bootstrap confidence intervals.}

\rev{Table~\ref{tab:liwc-domain-map} reports the theory-guided LIWC domains used in the interpretability analysis. Tables A12 and A13 report fixed-seed bootstrap confidence intervals for the domain-level Cohen's $d$ estimates shown in Figure~\ref{fig:liwc-domain-smd}. Table A14 reports estimates and confidence intervals for all LIWC categories included in the eight theory-guided domains. These results are descriptive and are not used to make cross-lingual psychometric equivalence claims.}

\FloatBarrier

\begin{table*}[!tbp]
\centering
\apptablesize
\setlength{\tabcolsep}{5pt}
\appstretch
\begin{tabularx}{\textwidth}{p{0.20\linewidth}XX}
\toprule
\textbf{LIWC domain} & \textbf{Categories used} & \textbf{Interpretive role} \\
\midrule
Self-focus & First-person singular; first-person plural & Self-reference and collective orientation \\
Social orientation & Social; affiliation; family; friend; other pronouns & Interpersonal orientation and social engagement \\
Negative affect & Negative emotion; sadness; anxiety; anger & Affective distress \\
Positive / reward orientation & Positive emotion; reward; leisure; achievement & Positive affect and reward-related orientation \\
Cognitive processing & Cognitive process; insight; causation; discrepancy; tentative; certainty; negation; comparison & Cognitive appraisal and evaluative processing \\
\rev{Somatic / biological} & \rev{Body; health; biological; illness; food/eating} & \rev{Somatic and biological concerns} \\
Temporal focus & Past focus; present focus; future focus; time & Temporal orientation \\
Risk / death & Death; risk & Mortality and threat-related language \\
\bottomrule
\end{tabularx}
\caption{Theory-guided LIWC domains used in the interpretability analysis. Category availability differs across LIWC versions and languages, so analyses are computed within dataset and interpreted descriptively.}
\label{tab:liwc-domain-map}
\end{table*}

\begin{table*}[t]
\centering
\footnotesize
\begin{tabular}{lllrr}
\toprule
Dataset & Domain & Cohen's $d$ [95\% CI] & Categories & Seed \\
\midrule
DAIC-WOZ & Self-focus & 0.10 [-0.11, 0.31] & 2 & 1687726375 \\
DAIC-WOZ & Social orientation & -0.17 [-0.35, 0.02] & 5 & 690062081 \\
DAIC-WOZ & Negative affect & 0.46* [0.25, 0.68] & 4 & 1620507241 \\
DAIC-WOZ & Positive / reward orientation & -0.21* [-0.37, -0.06] & 4 & 1241242460 \\
DAIC-WOZ & Cognitive processing & -0.08 [-0.23, 0.10] & 6 & 1203481592 \\
DAIC-WOZ & Somatic / biological & 0.15 [-0.04, 0.37] & 3 & 1410068254 \\
DAIC-WOZ & Temporal focus & 0.18* [0.05, 0.31] & 4 & 2168944131 \\
DAIC-WOZ & Risk / death & 0.24* [0.01, 0.48] & 2 & 2722476929 \\
E-DAIC & Self-focus & 0.15 [-0.05, 0.35] & 2 & 2675416336 \\
E-DAIC & Social orientation & 0.10 [-0.09, 0.28] & 5 & 670732549 \\
E-DAIC & Negative affect & 0.32* [0.14, 0.62] & 4 & 2217719840 \\
E-DAIC & Positive / reward orientation & -0.04 [-0.15, 0.09] & 4 & 3833156227 \\
E-DAIC & Cognitive processing & 0.06 [-0.05, 0.17] & 6 & 3481421112 \\
E-DAIC & Somatic / biological & 0.28* [0.09, 0.49] & 3 & 289219826 \\
E-DAIC & Temporal focus & -0.08 [-0.19, 0.03] & 4 & 2344997551 \\
E-DAIC & Risk / death & 0.19 [-0.02, 0.41] & 2 & 895272012 \\
EATD & Self-focus & -0.09 [-0.38, 0.23] & 2 & 3514312505 \\
EATD & Social orientation & 0.19 [-0.14, 0.50] & 5 & 1417092225 \\
EATD & Negative affect & -0.12 [-0.33, 0.10] & 4 & 3212471318 \\
EATD & Positive / reward orientation & -0.31* [-0.55, -0.10] & 4 & 1955131837 \\
EATD & Cognitive processing & -0.03 [-0.30, 0.24] & 8 & 3400549602 \\
EATD & Somatic / biological & 0.19 [-0.18, 0.61] & 3 & 4283162258 \\
EATD & Temporal focus & -0.15 [-0.35, 0.05] & 4 & 3860010083 \\
EATD & Risk / death & 0.28 [-0.16, 0.61] & 2 & 2745179976 \\
\bottomrule
\end{tabular}
\caption{\rev{Fixed-seed domain-level Cohen's $d$ estimates for DAIC-WOZ, E-DAIC, and EATD. Asterisks are descriptive, unadjusted, and indicate that the participant-bootstrap 95\% confidence interval excludes zero before rounding.}}
\label{tab:A12-domain-fixed-seed}
\end{table*}

\begin{table*}[t]
\centering
\footnotesize
\begin{tabular}{lllrr}
\toprule
Dataset & Domain & Cohen's $d$ [95\% CI] & Categories & Seed \\
\midrule
MODMA & Self-focus & -0.36 [-0.81, 0.05] & 2 & 2617333303 \\
MODMA & Social orientation & -0.34 [-0.74, 0.01] & 5 & 3139331698 \\
MODMA & Negative affect & 0.11 [-0.22, 0.72] & 4 & 2701272406 \\
MODMA & Positive / reward orientation & -0.51* [-0.98, -0.16] & 4 & 2223957007 \\
MODMA & Cognitive processing & -0.09 [-0.55, 0.31] & 8 & 3218436948 \\
MODMA & Somatic / biological & 0.14 [-0.13, 0.78] & 3 & 2900234461 \\
MODMA & Temporal focus & 0.05 [-0.43, 0.49] & 4 & 3718725768 \\
MODMA & Risk / death & -0.06 [-0.51, 0.42] & 2 & 3143077328 \\
PDCH & Self-focus & -0.29 [-0.77, 0.14] & 2 & 3746729913 \\
PDCH & Social orientation & -0.11 [-0.51, 0.23] & 5 & 1944736423 \\
PDCH & Negative affect & 0.32 [-0.11, 0.73] & 4 & 3125733377 \\
PDCH & Positive / reward orientation & 0.25 [-0.03, 0.54] & 4 & 2358885011 \\
PDCH & Cognitive processing & 0.22 [-0.03, 0.45] & 8 & 3311529296 \\
PDCH & Somatic / biological & -0.15 [-0.58, 0.27] & 3 & 3061871638 \\
PDCH & Temporal focus & 0.05 [-0.30, 0.42] & 4 & 1337896997 \\
PDCH & Risk / death & 0.05 [-0.37, 0.42] & 2 & 3838999029 \\
\bottomrule
\end{tabular}
\caption{\rev{Fixed-seed domain-level Cohen's $d$ estimates for MODMA and PDCH. Asterisks are descriptive, unadjusted, and indicate that the participant-bootstrap 95\% confidence interval excludes zero before rounding.}}
\label{tab:A13-domain-fixed-seed}
\end{table*}

\FloatBarrier

\clearpage
\onecolumn

\begingroup
\footnotesize
\begin{longtable}{llllrl}
\toprule
Dataset & Domain & Category & LIWC column & $d$ & 95\% CI \\
\midrule
DAIC-WOZ & Self-focus & First-person singular & i & 0.36 & [0.04, 0.68] \\
DAIC-WOZ & Self-focus & First-person plural & we & -0.17 & [-0.46, 0.15] \\
DAIC-WOZ & Social orientation & Social & Social & -0.06 & [-0.36, 0.27] \\
DAIC-WOZ & Social orientation & Affiliation & affiliation & -0.21 & [-0.50, 0.09] \\
DAIC-WOZ & Social orientation & Family & family & 0.07 & [-0.24, 0.40] \\
DAIC-WOZ & Social orientation & Friend & friend & -0.33 & [-0.61, -0.04] \\
DAIC-WOZ & Social orientation & Other pronouns & you & -0.30 & [-0.56, -0.03] \\
DAIC-WOZ & Negative affect & Negative emotion & emo\_neg & 0.49 & [0.17, 0.82] \\
DAIC-WOZ & Negative affect & Sadness & emo\_sad & 0.45 & [0.13, 0.81] \\
DAIC-WOZ & Negative affect & Anxiety & emo\_anx & 0.64 & [0.30, 1.01] \\
DAIC-WOZ & Negative affect & Anger & emo\_anger & 0.25 & [-0.13, 0.58] \\
DAIC-WOZ & Positive / reward orientation & Positive emotion & emo\_pos & -0.27 & [-0.59, 0.04] \\
DAIC-WOZ & Positive / reward orientation & Reward & reward & -0.42 & [-0.65, -0.17] \\
DAIC-WOZ & Positive / reward orientation & Leisure & leisure & -0.09 & [-0.38, 0.22] \\
DAIC-WOZ & Positive / reward orientation & Achievement & achieve & -0.08 & [-0.37, 0.23] \\
DAIC-WOZ & Cognitive processing & Cognitive process & cogproc & -0.15 & [-0.44, 0.12] \\
DAIC-WOZ & Cognitive processing & Insight & insight & -0.13 & [-0.40, 0.14] \\
DAIC-WOZ & Cognitive processing & Causation & cause & -0.15 & [-0.47, 0.16] \\
DAIC-WOZ & Cognitive processing & Discrepancy & discrep & 0.03 & [-0.32, 0.38] \\
DAIC-WOZ & Cognitive processing & Tentative & tentat & -0.24 & [-0.50, 0.02] \\
DAIC-WOZ & Cognitive processing & Negation & negate & 0.18 & [-0.08, 0.70] \\
DAIC-WOZ & Somatic / biological & Health & health & 0.49 & [0.19, 0.87] \\
DAIC-WOZ & Somatic / biological & Illness & illness & 0.07 & [-0.23, 0.40] \\
DAIC-WOZ & Somatic / biological & Food / eating & food & -0.10 & [-0.34, 0.19] \\
DAIC-WOZ & Temporal focus & Past focus & focuspast & 0.12 & [-0.20, 0.44] \\
DAIC-WOZ & Temporal focus & Present focus & focuspresent & 0.35 & [0.02, 0.70] \\
DAIC-WOZ & Temporal focus & Future focus & focusfuture & 0.04 & [-0.25, 0.34] \\
DAIC-WOZ & Temporal focus & Time & time & 0.19 & [-0.11, 0.50] \\
DAIC-WOZ & Risk / death & Death & death & 0.25 & [-0.08, 0.60] \\
DAIC-WOZ & Risk / death & Risk & risk & 0.22 & [-0.10, 0.55] \\
E-DAIC & Self-focus & First-person singular & i & 0.32 & [0.05, 0.59] \\
E-DAIC & Self-focus & First-person plural & we & -0.03 & [-0.27, 0.25] \\
E-DAIC & Social orientation & Social & Social & -0.00 & [-0.29, 0.28] \\
E-DAIC & Social orientation & Affiliation & affiliation & 0.11 & [-0.14, 0.39] \\
E-DAIC & Social orientation & Family & family & 0.35 & [0.07, 0.66] \\
E-DAIC & Social orientation & Friend & friend & 0.16 & [-0.22, 0.43] \\
E-DAIC & Social orientation & Other pronouns & you & -0.11 & [-0.40, 0.19] \\
E-DAIC & Negative affect & Negative emotion & emo\_neg & 0.37 & [0.09, 0.79] \\
E-DAIC & Negative affect & Sadness & emo\_sad & 0.10 & [-0.07, 0.63] \\
E-DAIC & Negative affect & Anxiety & emo\_anx & 0.60 & [0.27, 0.97] \\
E-DAIC & Negative affect & Anger & emo\_anger & 0.23 & [-0.03, 0.65] \\
E-DAIC & Positive / reward orientation & Positive emotion & emo\_pos & -0.13 & [-0.30, 0.08] \\
E-DAIC & Positive / reward orientation & Reward & reward & -0.15 & [-0.37, 0.10] \\
E-DAIC & Positive / reward orientation & Leisure & leisure & -0.10 & [-0.27, 0.10] \\
E-DAIC & Positive / reward orientation & Achievement & achieve & 0.24 & [-0.02, 0.52] \\
E-DAIC & Cognitive processing & Cognitive process & cogproc & 0.18 & [-0.04, 0.39] \\
E-DAIC & Cognitive processing & Insight & insight & -0.01 & [-0.21, 0.27] \\
E-DAIC & Cognitive processing & Causation & cause & -0.02 & [-0.22, 0.24] \\
E-DAIC & Cognitive processing & Discrepancy & discrep & 0.03 & [-0.17, 0.29] \\
E-DAIC & Cognitive processing & Tentative & tentat & 0.08 & [-0.14, 0.32] \\
E-DAIC & Cognitive processing & Negation & negate & 0.09 & [-0.20, 0.38] \\
E-DAIC & Somatic / biological & Health & health & 0.54 & [0.26, 0.85] \\
E-DAIC & Somatic / biological & Illness & illness & 0.22 & [-0.07, 0.55] \\
E-DAIC & Somatic / biological & Food / eating & food & 0.08 & [-0.14, 0.35] \\
E-DAIC & Temporal focus & Past focus & focuspast & 0.14 & [-0.13, 0.41] \\
E-DAIC & Temporal focus & Present focus & focuspresent & -0.21 & [-0.40, -0.02] \\
E-DAIC & Temporal focus & Future focus & focusfuture & -0.17 & [-0.30, 0.01] \\
E-DAIC & Temporal focus & Time & time & -0.09 & [-0.26, 0.11] \\
E-DAIC & Risk / death & Death & death & 0.24 & [-0.09, 0.55] \\
E-DAIC & Risk / death & Risk & risk & 0.14 & [-0.11, 0.46] \\
EATD & Self-focus & First-person singular & i & -0.06 & [-0.49, 0.37] \\
EATD & Self-focus & First-person plural & we & -0.12 & [-0.37, 0.22] \\
EATD & Social orientation & Social & social & 0.20 & [-0.31, 0.71] \\
EATD & Social orientation & Affiliation & affiliation & -0.08 & [-0.59, 0.41] \\
EATD & Social orientation & Family & family & 0.58 & [0.00, 1.07] \\
EATD & Social orientation & Friend & friend & 0.06 & [-0.42, 0.55] \\
EATD & Social orientation & Other pronouns & you & 0.21 & [-0.31, 0.73] \\
EATD & Negative affect & Negative emotion & negemo & 0.05 & [-0.35, 0.49] \\
EATD & Negative affect & Sadness & sad & -0.17 & [-0.44, 0.18] \\
EATD & Negative affect & Anxiety & anx & -0.24 & [-0.49, 0.06] \\
EATD & Negative affect & Anger & anger & -0.11 & [-0.50, 0.32] \\
EATD & Positive / reward orientation & Positive emotion & posemo & -0.28 & [-0.71, 0.15] \\
EATD & Positive / reward orientation & Reward & reward & -0.24 & [-0.58, 0.15] \\
EATD & Positive / reward orientation & Leisure & leisure & -0.35 & [-0.62, -0.07] \\
EATD & Positive / reward orientation & Achievement & achieve & -0.39 & [-0.79, 0.02] \\
EATD & Cognitive processing & Cognitive process & cogproc & -0.04 & [-0.53, 0.45] \\
EATD & Cognitive processing & Insight & insight & 0.15 & [-0.33, 0.65] \\
EATD & Cognitive processing & Causation & cause & -0.13 & [-0.49, 0.24] \\
EATD & Cognitive processing & Discrepancy & discrep & -0.30 & [-0.72, 0.11] \\
EATD & Cognitive processing & Tentative & tentat & -0.31 & [-0.70, 0.07] \\
EATD & Cognitive processing & Certainty & certain & 0.12 & [-0.29, 0.58] \\
EATD & Cognitive processing & Negation & negate & 0.63 & [0.16, 1.16] \\
EATD & Cognitive processing & Comparison & compare & -0.34 & [-0.77, 0.09] \\
EATD & Somatic / biological & Body & body & 0.30 & [-0.17, 0.82] \\
EATD & Somatic / biological & Health & health & -0.15 & [-0.43, 0.19] \\
EATD & Somatic / biological & Biological & bio & 0.44 & [-0.02, 0.93] \\
EATD & Temporal focus & Past focus & focuspast & -0.02 & [-0.42, 0.41] \\
EATD & Temporal focus & Present focus & focuspresent & 0.07 & [-0.26, 0.56] \\
EATD & Temporal focus & Future focus & focusfuture & -0.37 & [-0.71, -0.01] \\
EATD & Temporal focus & Time & time & -0.30 & [-0.72, 0.12] \\
EATD & Risk / death & Death & death & 0.43 & [NA, NA] \\
EATD & Risk / death & Risk & risk & 0.13 & [-0.21, 0.57] \\
MODMA & Self-focus & First-person singular & i & -0.18 & [-0.82, 0.36] \\
MODMA & Self-focus & First-person plural & we & -0.54 & [-1.00, -0.06] \\
MODMA & Social orientation & Social & social & -0.49 & [-1.22, 0.07] \\
MODMA & Social orientation & Affiliation & affiliation & -0.69 & [-1.24, -0.19] \\
MODMA & Social orientation & Family & family & -0.26 & [-0.88, 0.30] \\
MODMA & Social orientation & Friend & friend & -0.48 & [-1.07, 0.03] \\
MODMA & Social orientation & Other pronouns & you & 0.20 & [-0.35, 0.75] \\
MODMA & Negative affect & Negative emotion & negemo & 0.21 & [-0.24, 1.23] \\
MODMA & Negative affect & Sadness & sad & -0.16 & [-0.45, 0.73] \\
MODMA & Negative affect & Anxiety & anx & 0.14 & [-0.42, 0.70] \\
MODMA & Negative affect & Anger & anger & 0.25 & [-0.31, 0.89] \\
MODMA & Positive / reward orientation & Positive emotion & posemo & -0.53 & [-1.11, 0.02] \\
MODMA & Positive / reward orientation & Reward & reward & -0.56 & [-1.17, -0.03] \\
MODMA & Positive / reward orientation & Leisure & leisure & -0.39 & [-0.91, -0.09] \\
MODMA & Positive / reward orientation & Achievement & achieve & -0.59 & [-1.24, -0.05] \\
MODMA & Cognitive processing & Cognitive process & cogproc & -0.14 & [-0.77, 0.42] \\
MODMA & Cognitive processing & Insight & insight & 0.36 & [-0.23, 0.79] \\
MODMA & Cognitive processing & Causation & cause & -0.49 & [-1.03, 0.02] \\
MODMA & Cognitive processing & Discrepancy & discrep & -0.67 & [-1.33, -0.12] \\
MODMA & Cognitive processing & Tentative & tentat & -0.58 & [-1.26, -0.03] \\
MODMA & Cognitive processing & Certainty & certain & 0.09 & [-0.50, 0.64] \\
MODMA & Cognitive processing & Negation & negate & 0.98 & [0.50, 1.57] \\
MODMA & Cognitive processing & Comparison & compare & -0.31 & [-0.94, 0.28] \\
MODMA & Somatic / biological & Body & body & 0.70 & [0.15, 1.38] \\
MODMA & Somatic / biological & Health & health & -0.10 & [-0.66, 0.47] \\
MODMA & Somatic / biological & Biological & bio & -0.18 & [-0.42, 1.18] \\
MODMA & Temporal focus & Past focus & focuspast & 0.41 & [-0.17, 0.95] \\
MODMA & Temporal focus & Present focus & focuspresent & 0.44 & [-0.13, 1.01] \\
MODMA & Temporal focus & Future focus & focusfuture & -0.61 & [-1.29, -0.06] \\
MODMA & Temporal focus & Time & time & -0.03 & [-0.63, 0.53] \\
MODMA & Risk / death & Death & death & -0.19 & [-0.72, 0.37] \\
MODMA & Risk / death & Risk & risk & 0.08 & [-0.43, 0.72] \\
PDCH & Self-focus & First-person singular & i & 0.02 & [-0.35, 0.40] \\
PDCH & Self-focus & First-person plural & we & -0.59 & [-1.35, 0.07] \\
PDCH & Social orientation & Social & social & -0.17 & [-0.67, 0.29] \\
PDCH & Social orientation & Affiliation & affiliation & -0.27 & [-0.95, 0.34] \\
PDCH & Social orientation & Family & family & 0.01 & [-0.54, 0.46] \\
PDCH & Social orientation & Friend & friend & 0.32 & [-0.24, 0.84] \\
PDCH & Social orientation & Other pronouns & you & -0.47 & [-1.21, 0.17] \\
PDCH & Negative affect & Negative emotion & negemo & 0.26 & [-0.34, 0.85] \\
PDCH & Negative affect & Sadness & sad & 0.23 & [-0.38, 0.84] \\
PDCH & Negative affect & Anxiety & anx & 0.21 & [-0.32, 0.72] \\
PDCH & Negative affect & Anger & anger & 0.55 & [0.06, 1.06] \\
PDCH & Positive / reward orientation & Positive emotion & posemo & 0.49 & [0.03, 1.01] \\
PDCH & Positive / reward orientation & Reward & reward & -0.21 & [-0.70, 0.22] \\
PDCH & Positive / reward orientation & Leisure & leisure & 0.34 & [-0.21, 0.86] \\
PDCH & Positive / reward orientation & Achievement & achieve & 0.38 & [-0.20, 0.97] \\
PDCH & Cognitive processing & Cognitive process & cogproc & 0.39 & [-0.14, 0.92] \\
PDCH & Cognitive processing & Insight & insight & -0.28 & [-0.85, 0.29] \\
PDCH & Cognitive processing & Causation & cause & 0.64 & [0.18, 1.11] \\
PDCH & Cognitive processing & Discrepancy & discrep & 0.46 & [-0.05, 0.96] \\
PDCH & Cognitive processing & Tentative & tentat & 0.32 & [-0.26, 0.88] \\
PDCH & Cognitive processing & Certainty & certain & 0.13 & [-0.32, 0.54] \\
PDCH & Cognitive processing & Negation & negate & 0.15 & [-0.32, 0.55] \\
PDCH & Cognitive processing & Comparison & compare & -0.08 & [-0.67, 0.40] \\
PDCH & Somatic / biological & Body & body & 0.24 & [-0.21, 0.65] \\
PDCH & Somatic / biological & Health & health & -0.42 & [-1.03, 0.15] \\
PDCH & Somatic / biological & Biological & bio & -0.26 & [-0.86, 0.28] \\
PDCH & Temporal focus & Past focus & focuspast & 0.18 & [-0.46, 0.77] \\
PDCH & Temporal focus & Present focus & focuspresent & -0.18 & [-0.72, 0.30] \\
PDCH & Temporal focus & Future focus & focusfuture & 0.16 & [-0.47, 0.79] \\
PDCH & Temporal focus & Time & time & 0.04 & [-0.59, 0.65] \\
PDCH & Risk / death & Death & death & 0.14 & [-0.54, 0.74] \\
PDCH & Risk / death & Risk & risk & -0.04 & [-0.56, 0.35] \\
\bottomrule
\\[1ex]
\caption{Category-level fixed-seed Cohen's d estimates and stratified participant-bootstrap percentile intervals. Deterministic seed specifications are retained in the released analysis artifacts. The confidence interval for the EATD Death category is reported as [NA, NA] because the category was extremely sparse: only 1 of 30 participants with depression and 0 of 132 participants without depression had nonzero values, causing many bootstrap resamples to have zero pooled variance.}
\label{tab:A14-category-fixed-seed}
\end{longtable}
\endgroup

\end{document}